# Magnetic and electrical properties of $RCo_2Mn$ (R=Ho, Er) compounds

Bibekananda Maji[1], K. G. Suresh[1,*] and A. K. Nigam[2]
[1]Magnetic Materials Laboratory, Department of Physics,
Indian Institute of Technology Bombay, Mumbai- 400076, India
[2]Tata Institute of Fundamental Research,
Homi Bhabha Road, Mumbai- 400005, India

## Abstract

The $RCo_2Mn$ (R= Ho and Er) alloys, crystallizing in the cubic $MgCu_2$-type structure, are isostructural to $RCo_2$ compounds. The excess Mn occupies both the R and the Co atomic positions. Magnetic, electrical and heat capacity measurements have been done in these comounds**.** The Curie temperature is found to be 248 K and 222 K for $HoCo_2Mn$ and $ErCo_2Mn$ respectively, which are considerably higher than that of the corresponding $RCo_2$ compounds. Saturation magnetization values calculated in these samples are less compared to that of the corresponding $RCo_2$ compounds. Heat capacity data have been fitted with the nonmagnetic contribution $(\gamma T + C_{latt})$ with $\theta_D$ =250 K and $\gamma$ =26 mJ $mol^{-1}K^{-2}$.

*Corresponding author (email: suresh@phy.iitb.ac.in)

## 1.Introduction

The study of rare earth(R)-transition metal (T) intermetallic compounds has led to the discovery of various novel materials, some of them displaying interesting physical properties such as giant megnetoresistance, giant magnetostriction, giant magnetocaloric effect etc. [1]. It is observed that usually for rare-earth transition metal intermetallic compounds, the 4f-3d spin-spin coupling is antiferromagnetic, leading to a parallel alignment of 3d and 4f moments in the light lanthanide compounds and to an antiparallel alignment in the heavy lanthanide compounds [2, 3]. Among the R-T compounds, the cubic Laves-phase $RCo_2$ compounds attracted considerable attention from researchers due to their special features arising from the interaction of the well localized rare-earth 4f electrons with the itinerant 3d electrons of Co [2, 3, 4]. In this series, the compounds with R= Er, Ho, and Dy have been subjected to intense research due to the occurrence of first order magnetic phase transition at their Curie temperature ($T_C$). The first order transition is due to the itinerant electron metamagnetism occurring in the Co sublattice. The exchange field of ordered 4f moments in these three compounds exceeds the critical field required for the metamagnetic transition of Co 3d electrons, which causes an abrupt increase in the Co magnetic moment from zero to about 0.7-1.0 $\mu_B$ via a first order phase transition [4] .

Recently it was reported that Mn can be substituted in $RNi_2$ series, which results in the modification of the magnetic properties considerably [5]. Wang et al. reported that similar to $RNi_2$ alloys, the $RNi_2Mn$ with R= Tb, Dy, Ho, Er alloys also crystallize in the cubic $MgCu_2$- type of structure with the Mn atoms occupying both the R and the Ni lattice positions [5]. Their Curie temperatures were found to be considerably higher than that of the corresponding $RNi_2$ compounds whereas the spontaneous magnetic moments are smaller than that of the corresponding $RNi_2$ alloys [5, 6]. In the $RNi_2Mn$ alloys, it is assumed that the Ni atoms have a negligible magnetic moment whereas Mn atoms possess a large magnetic moment directed opposite to that of the R atoms. [5, 6]. Considering the similarities between $RNi_2$ and $RCo_2$ compounds, it is of importance to study such an influence of Mn on the latter compounds as well. Therefore, we have studied $RCo_2Mn$ (with R= Ho, Er) compounds which are isostructural to $RCo_2$ compounds. In this paper, we report the structural, magnetic, transport and thermal properties of these compounds,

## 2. Experimental details

$RCo_2Mn$ (R= Ho, Er) compounds were synthesized by arc melting the starting elements with purity of at least 99.9%, on water cooled copper hearth in argon atmosphere. The constituents were melted in the stoichiometric amounts 1:2:1. Excess of Mn (of about 6%) was added to compensate the evaporation during the melting. The resulting ingots were turned upside down and remelted several times to ensure homogeneity. Subsequently, they were annealed in high-purity argon atmosphere at 800 $^{o}$C for a week. The structural analysis of the samples was performed by collecting the room temperature powder x-ray diffractograms (XRD) using Cu-K$\alpha$ radiation. The magnetization measurements were carried out using a vibrating sample magnetometer attached to a Physical Property Measurement System (Quantum Design, Design, PPMS-6500). The resistivity measurements were done using pour probe technique and using a PPMS. The measurement of the heat capacity was performed using the thermal relaxation technique in the PPMS.

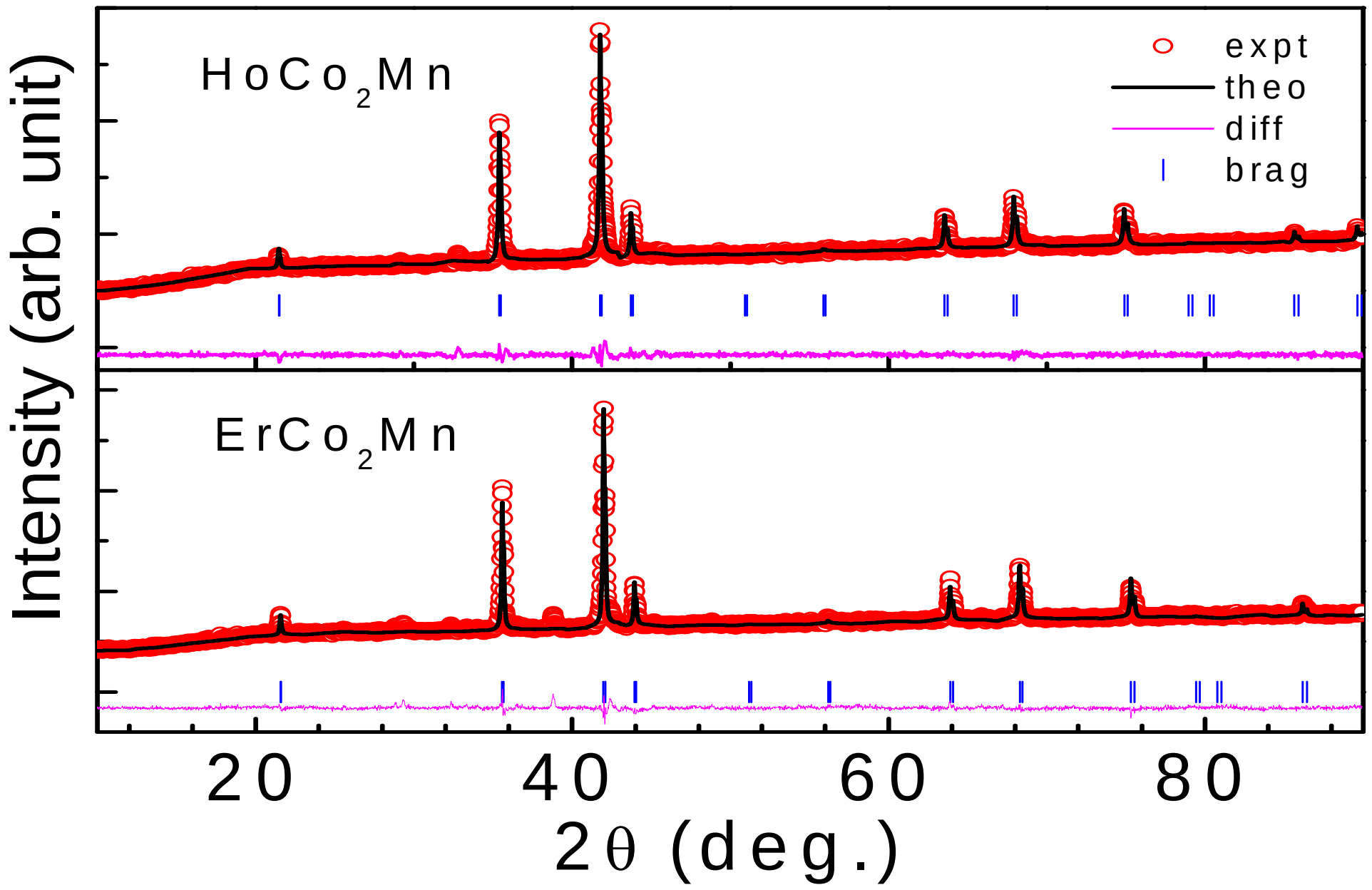

FIG. 1. Room temperature x-ray diffraction patterns, along with the Rietveld refined patterns of $RCo_2Mn$ compounds. The plots at the bottom show the difference between theoretical and experimental data.

## 3. Results and Discussion

Fig.1 shows the room temperature powder x-ray diffraction patterns of $RCo_2Mn$ compounds. The Rietveld refined plot is shown for the compounds in each case. The difference plot between the theoretical and the experimentally observed patterns is shown at the bottom of the figure. The refinement confirms that both these compounds are single phase, possessing the $MgCu_2$ type C-15 cubic Laves phase structure (space group Fd3m), except for traces of impurities. The site occupancy in $HoCo_2Mn$ is listed in Table 1. Although the chemical composition of the compounds is $RCo_2Mn$, its crystallographic structure is similar to that of $RCo_2$ in which *R* and *Co* atoms occupy the 8*a* and 16*d* sites respectively. The same trend is seen between $RNi_2Mn$ and $RNi_2$ compounds [5]. In Laves phase structure, the 8a sites are not completely occupied by R atoms and some Mn atoms occupy these vacancies [5,7]. As can be seen from Table 1, in the case of $HoCo_2Mn$, the best fitting is obtained when about 22.1 % of 8a sites are occupied by Mn and 3.8 % sites are empty. A similar site occupancy was observed for $ErCo_2Mn$ as well. The variation of the lattice parameters between $RCo_2$ and $RCo_2Mn$ compounds are given in Table II.

TABLE I. Rietveld refinement results for $HoCo_2Mn$.

| Atom | Positions | x/a | y/a | z/a | Site occupancy |
|---|---|---|---|---|---|
| Ho | 8a | 0 | 0 | 0 | 75.1% |
| Vacancy | 8a | 0 | 0 | 0 | 3.8% |
| Mn | 8a | 0 | 0 | 0 | 22.1% |
| Co | 16d | 0.625 | 0.625 | 0.625 | 75.1% |
| Mn | 16d | 0.625 | 0.625 | 0.625 | 24.9% |

TABLE II. Lattice parameters, Curie temperature and saturation magnetization of $RCo_2Mn$ compounds along with those of $RCo_2$ compounds (error in lattice parameter is $\pm 0.001 \overset{o}{A}$).

| Compound | a ($\overset{o}{A}$) | $T_c$ (K) | $M_s$ ($\mu_B / f.u$) | Ref. |
|---|---|---|---|---|
| $HoCo_2Mn$ | 7.166 | 248 | 6.6 | This work |
| $ErCo_2Mn$ | 7.132 | 221 | 5.2 | This work |
| $HoCo_2$ | 7.151 | 78 | 7.8 | [8,10] |
| $ErCo_2$ | 7.153 | 35 | 6.0 | [9,10] |

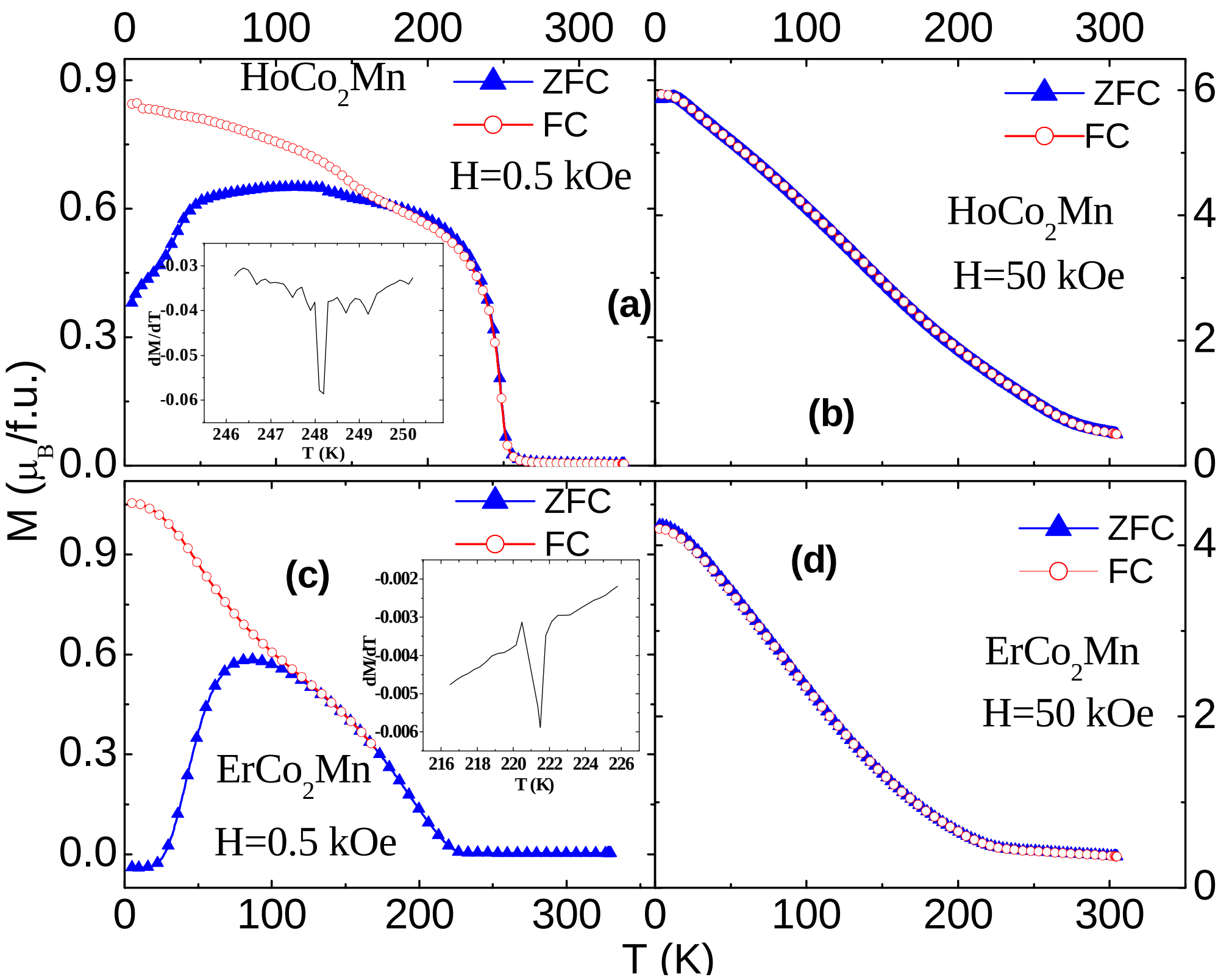

FIG. 2. Temperature dependence of magnetization of $HoCo_2Mn$ in fields of 0.5 kOe and 50 kOe (a and b) and $ErCo_2Mn$ (c and d). The insets in (a) and (c) show temperature dependency of the first derivative of magnetization.

The temperature (T) variation of magnetization (M) data has been collected in applied fields (H) of 0.5 and 50 kOe for the $RCo_2Mn$ compounds, under both the zero-field-cooled (ZFC) and field-cooled warming (FCW) conditions. A large thermomagnetic irreversibility between the ZFC and FCW data is found at 0.5 kOe in the temperature range of 3-110 K for $ErCo_2Mn$ and 3-160 K for $HoCo_2Mn$. The thermomagnetic irreversibility is found to decrease with increasing applied field and it vanishes completely at 50 KOe. The irreversible temperature dependences of the magnetization as displayed in figure can be considered as a magnetohistory effect resulting from the presence of narrow domain walls [5, 7, 11, 12] or as cluster glass behavior [5, 7, 13 ].

The $T_C$ values of these compounds determined from the first derivative (*dM/dT*) of the data at 500 Oe are listed in Table II. It is found that the values are 248 and 221 K for $HoCo_2Mn$ and $ErCo_2Mn$, respectively. These values are much higher than that of the corresponding $RCo_2$ and $RMn_2$ compounds, implying that the exchange interaction in $RCo_2Mn$ compounds are stronger than that in $RCo_2$ compounds [5]. It may be noted from Fig 2 that there is no visible anomaly corresponding to the Curie temperatures of the parent compounds namely $HoCo_2$ and $ErCo_2$.

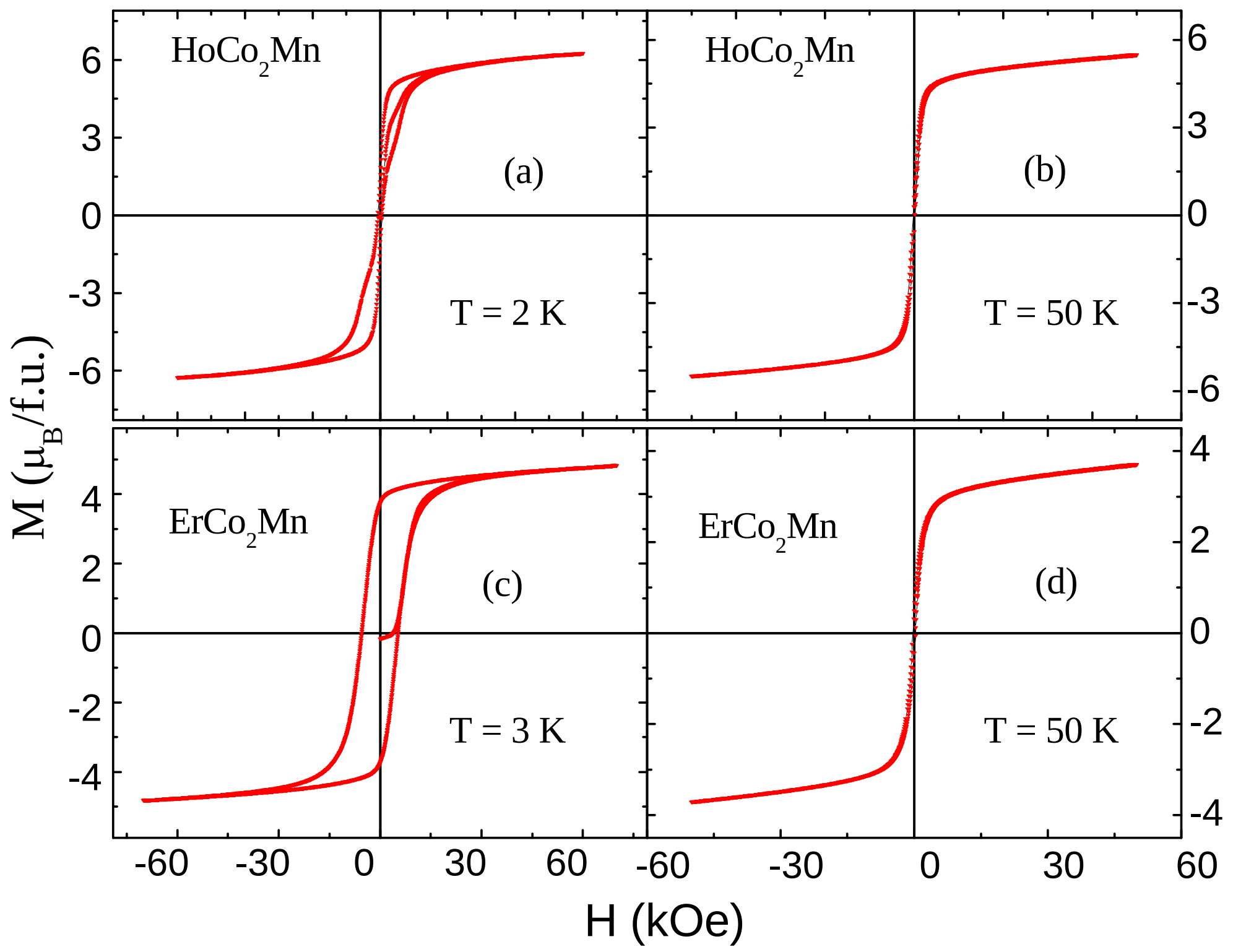

FIG. 3. (a, b) Hysteresis loops of $HoCo_2Mn$ at 2 K and 50 K (a and b) and $ErCo_2Mn$ at 3 K and 50 K (c,d)

The magnetic hysteresis loop measurement is a useful and effective tool to study the magnetic domain dynamics [5]. Therefore, the hysteresis loops of $HoCo_2Mn$ and $ErCo_2Mn$ compounds have been recorded at low temperatures as well as at 50 K. From the hysteresis loops, the coercive field ($H_c$) of $ErCo_2Mn$ is found to be 5.4 kOe at 3*K* while it is 0.6 kOe for $HoCo_2Mn$ at 2K. This observation is in agreement with the larger thermomagnetic irreversibility seen in $ErCo_2Mn$ as compared to that in $HoCo_2Mn$. With increase in temperature, the coercivity decreases in both the compounds. The saturation magnetizations are 6.6 $\mu_B / f.u$ for $HoCo_2Mn$ and 5.2 $\mu_B / f.u$ for $ErCo_2Mn$. Both these values are less than that of the corresponding $RCo_2$ compounds.

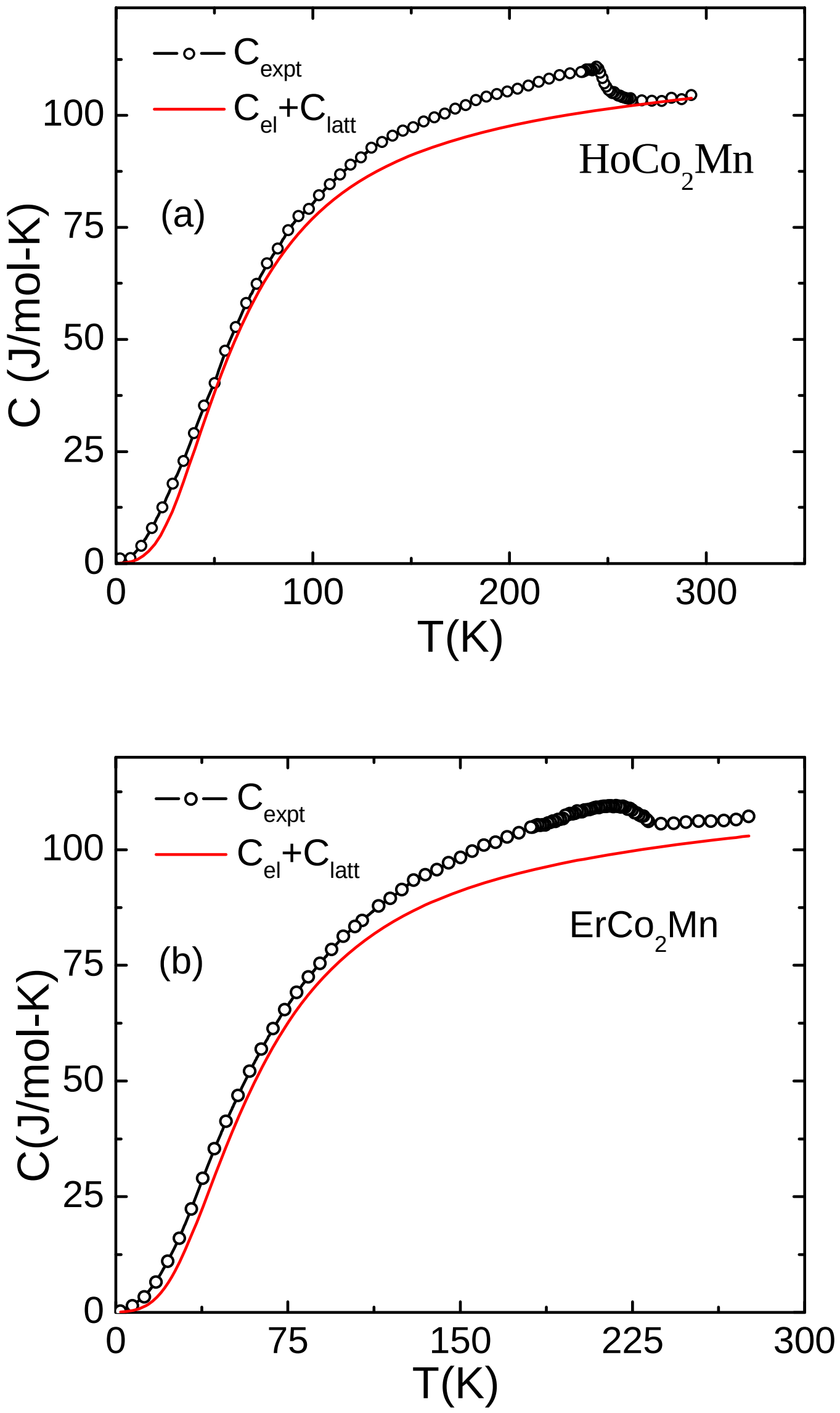

FIG. 4. Temperature dependence of the heat capacity at zero field (open circles) for (a) $HoCo_2Mn$ and (b) $ErCo_2Mn$. The solid lines represent the theoretically calculated electronic and lattice contributions ($C_{el}+C_{latt}$).

The variation of heat capacity (C) as a function of temperature for $HoCo_2Mn$ and $ErCo_2Mn$ at H=0 are shown in Fig. 4.(a, b). The zero field heat capacity shows a peak for both the samples, which nearly coincides with the $T_C$ values obtained from the temperature dependence of magnetization data. From the shape of the peaks observed in the C-T data, one can conclude that the magnetic transition is second in both these compounds. Heat capacity data also shows that there is no anomaly corresponding to the Curie temperatures of $HoCo_2$ and $ErCo_2$. A preliminary analysis of the C-T dependence has been carried out by separating various contributions to the total heat capacity. The total heat capacity can be written as sum of the electronic, lattice and magnetic contributions:

$$C_{tot} = C_{el} + C_{latt} + C_M = \gamma T + 9NR(T/\theta_D)^3 \int_0^{\theta_D/T} \frac{x^4 e^x}{(e^x - 1)^2} dx + C_m$$

Where N is the number of atoms per formula unit, R is the molar gas constant, $\gamma$ is the electronic coefficient and $\theta_D$, the Debye temperature. The sum of the lattice and electronic contributions $(\gamma T + C_{latt})$ can be obtained from the data of isostructural nonmagnetic compounds $YCo_2$ and $LuCo_2$. The values of $\gamma$ reported in the literature vary from 34 to 36.5 $mJmol^{-1}K^{-2}$ for $YCo_2$ and 26.7 to 36.5 $mJmol^{-1}K^{-2}$ for $LuCo_2$. The values reported for $\theta_D$ are in the range of 226-305 K for $YCo_2$ and 238-280 K for $LuCo_2$ [4, 14-18]. By using the values of $\theta_D$ =250 K and $\gamma$ =26 mJ $mol^{-1}K^{-2}$, we have calculated the nonmagnetic contribution to the heat capacity in the present case and a reasonable agreement has been ontained for the low temperature ($T<T_C$) regime. The fit is shown by solid lines in Fig. 4. (a) and (b).

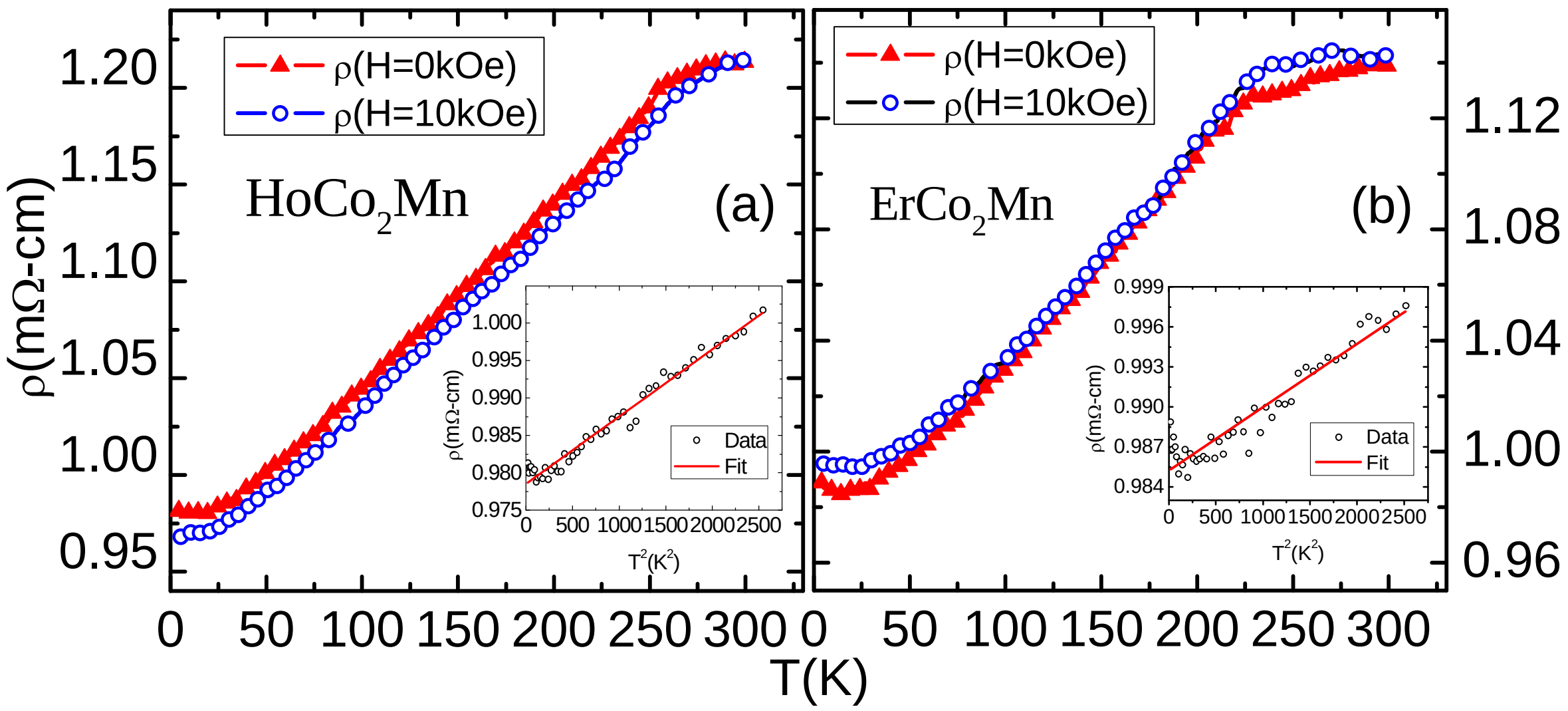

FIG. 5. The temperature dependence of electrical resistivity of (a) $HoCo_2Mn$ and (b) $ErCo_2Mn$ at H=0 and 50 kOe. The inset shows the fitting of the zero field data with the equation $\rho(T) = \rho_0 + AT^2$, over the temperature range of 5 - 50 K.

Fig.5.shows the temperature variation of electrical resistivity ($\rho$) of $HoCo_2Mn$ and $ErCo_2Mn$ with H=0 and 50 kOe The samples were first cooled in zero field and then data were recorded during warming cycle with applying field ( 0 or 50 kOe). As can be seen, both the compounds show metallic nature in the entire temperature range and show a slope change around the Curie temperature. However, the residual resistivity is found to be two orders of magnitude higher than that of the $RCo_2$ compounds [21, 22]. According to Mattheissen's rule , the total resistivity of a ferromagnetic material can be written as

$$\rho(\boldsymbol{T}) = \rho_0 + \rho_{ph}(\boldsymbol{T}) + \rho_{mag}(\boldsymbol{T}), \tag{1}$$

Where $\rho_0$ is the residual resistivity, $\rho_{ph}(\boldsymbol{T})$ is the contribution from the electron-phonon interaction, and $\rho_{mag}(\boldsymbol{T})$ is the contribution from electron-spin wave scattering. Depending on the temperature region, either $\rho_{ph}(\boldsymbol{T})$ or $\rho_{mag}(\boldsymbol{T})$ predominates. The temperature variations of the electrical resistivity due to the electron-electron and electron-spin wave scatterings follow a $\boldsymbol{T^2}$ behavior [23]. The resistivity in both the compounds is found to obey the relation $\rho(T) = \rho_0 + AT^2$ over the temperature range 5-50 K with $\rho_0$=0.978523 $m\Omega - cm$ and A=8.9682 $n\Omega - cm/k^2$ for $HoCo_2Mn$ and $\rho_0$=0.985242 $m\Omega - cm$ and 4.74645 $n\Omega - cm/k^2$ for $ErCo_2Mn$. From the values of the coefficient A, it is inferred that, in this temperature range, the electron-magnon scattering is dominant. At higher temperatures, electron-phonon interaction is found to be dominant, which gives a linear increase in the resistivity with temperature. The magnetoresistance calculated from the zero field and 10 kOe field data shows that the maximum values are 1.24% for $HoCo_2Mn$ at 5K and +0.99% for ErCo2Mn at 14K.

## **4.** Conclusions

We successfully synthesized novel compounds $RCo_2Mn$ with R= Ho, Er which crystallize in the cubic Laves-phase structure ($MgCu_2$-type, space group $\boldsymbol{Fd\bar{3}m}$) at room temperature. These compounds belong to a new family of intermetallic R-T compounds with interesting magnetic phenomena. In these compounds Mn atoms give a large magnetic moment opposite to the R atoms which leads to the strong R-Mn exchange coupling. As a result the Curie temperatures in these compounds are large compared to the corresponding $RCo_2$ compounds. The magnetic, transport and heat capacity data show that there are no anomalies corresponding to the ordering temperature of the parent $RCo_2$ compounds. The maximum magnetoresistance values are 1.24% for $HoCo_2Mn$ at 5K and +0.99% for ErCo2Mn at 14K for a field change of 10 kOe.